# When Students Write with AI: Lessons from Power Electronics Education

***Min Yang, Fanfan Lin***
(contributed equally)

## Abstract
Generative AI has quickly become a routine writing aid for power electronics engineers and students. Based on our teaching and mentoring experience in power electronics and academic writing, however, we have observed two recurring problems in student technical writing: compromised technical accuracy and declining readability. More importantly, students may lack the technical expertise and writing judgment needed to recognize and correct these issues. This article examines representative, anonymized student writing samples, including undergraduate theses and research papers in power electronics, to illustrate errors and weaknesses that students themselves failed to identify. Through these real-world examples, we discuss how students and educators can adapt technical writing practices to develop the domain expertise and critical judgment needed in the era of AI.

## I. Introduction

Generative artificial intelligence (GenAI) has become a routine part of technical writing and publishing. One important benefit is its potential to act as a linguistic equalizer in global science by lowering language barriers, particularly for writers from non-English-speaking backgrounds [1], [2], [3]. At the same time, however, we have observed an increasingly common mindset among engineering students: as long as the technical contribution is their "real work," the writing can be outsourced to AI. But is this shortcut as trustworthy as it appears?

From our teaching and mentoring experience in power electronics research and academic writing, we have repeatedly seen inexperienced writers encounter problems when relying on GenAI to prepare papers and theses. More concerning is that they often struggle to recognize these problems themselves due to lacking the domain expertise and writing judgment.

This article focuses on two recurring problems: technical inaccuracy and declining readability. A general-purpose GenAI tool, not customized for the power electronics domain or for high-quality technical writing, may not have the nuanced understanding to ensure that technical information presented is accurate and up to date. Hallucination in large language models remains an inherent concern. Texts generated or polished by AI tools can become unnecessarily long, formulaic, and semantically vague. We examine readability through three aspects particularly relevant to technical communication: word and sentence length, lexical and syntactic complexity, and lack of clarity.

To illustrate these issues, we examine representative, anonymized AI-assisted student writing samples from undergraduate theses and research manuscripts in power electronics. Through these real cases, we identify recurring problems that students themselves failed to recognize and offer practical reflections for both students and educators on how technical writing and writing instruction should adapt in the era of AI.

## II. When Fluent Writing is Technically Wrong

One of the most concerning problems with AI-assisted technical writing is that incorrect content may not appear obviously wrong to a non-expert. Instead, it can look fluent, professional, and technically plausible even when it contains factual errors and nonstandard terminology. This problem is not limited to text; it can also appear in AI-generated figures. The following examples show how AI-generated content can appear convincing at first glance while still requiring solid technical judgment to verify.

***Sample 1 (Excerpt from an undergraduate thesis on a dual-active-bridge converter)***

> *"All efficiencies are simulated power-stage efficiencies; front-end bridge, EMI, protection, and wall-plug losses are outside the circuit-simulation model unless explicitly stated."*

The student attempted to clarify which losses were included in the converter simulation and which were excluded. To a student new to power electronics, the sentence may appear professional because it combines several familiar technical terms. A closer examination, however, reveals multiple technical problems.

First, "front-end bridge" is not a standard term for the DAB converter; the topology is normally described by its primary- and secondary-side full bridges. Second, both full bridges are integral parts of the DAB power stage and contribute to conduction and switching losses, so it would not be acceptable practice to leave out one of the bridges for efficiency analysis. Third, "front-end bridge, EMI, protection, and wall-plug losses" refer to concepts at different system levels and should not be listed as parallel loss categories. EMI refers to electromagnetic interference, protection is a system function, and wall-plug loss describes a system-level efficiency boundary.

To avoid such problems, students should not outsource underlying technical reasoning to GenAI. They should first understand the relevant concepts and establish a technically sound loss breakdown before using GenAI to help formulate the text.

**Sample 2** ***(Excerpt from an undergraduate thesis on converter design optimization)***

> *"Moreover, different applications may require different trade-offs: safety-critical systems usually demand higher ZVS reliability, whereas charging or energy-conversion systems may place more emphasis on reducing current stress."*

In this sentence, the student attempted to articulate that power converters used in different applications may emphasize different performance metrics. The statement reads fluently and sounds plausible, yet it contains both imprecise terminology and a flawed comparison.

In power electronics, “ZVS reliability” is not a commonly used technical term. ZVS refers to the realization of zero-voltage switching at the intended switching transitions. Reliability refers to the ability of power electronic components or systems to perform their intended functions consistently over time. These are fundamentally different concepts. More importantly, the comparison itself is conceptually flawed. “Safety-critical systems” describes the criticality or reliability requirements of a system, whereas “charging” and “energy-conversion systems” describe application or functional categories. These categories are not mutually exclusive; in fact, charging systems are themselves energy-conversion systems and may also be safety-critical. Therefore, they should not be presented as opposing application categories in this sentence.

A possible revision is: *“Different applications prioritize different objectives. Safety-critical applications such as electric aircraft and medical power supplies place greater emphasis on reliability-related metrics, whereas consumer-electronics converters prioritize cost and power density.”*

**Sample 3 (AI-generated figure used in an undergraduate thesis)**

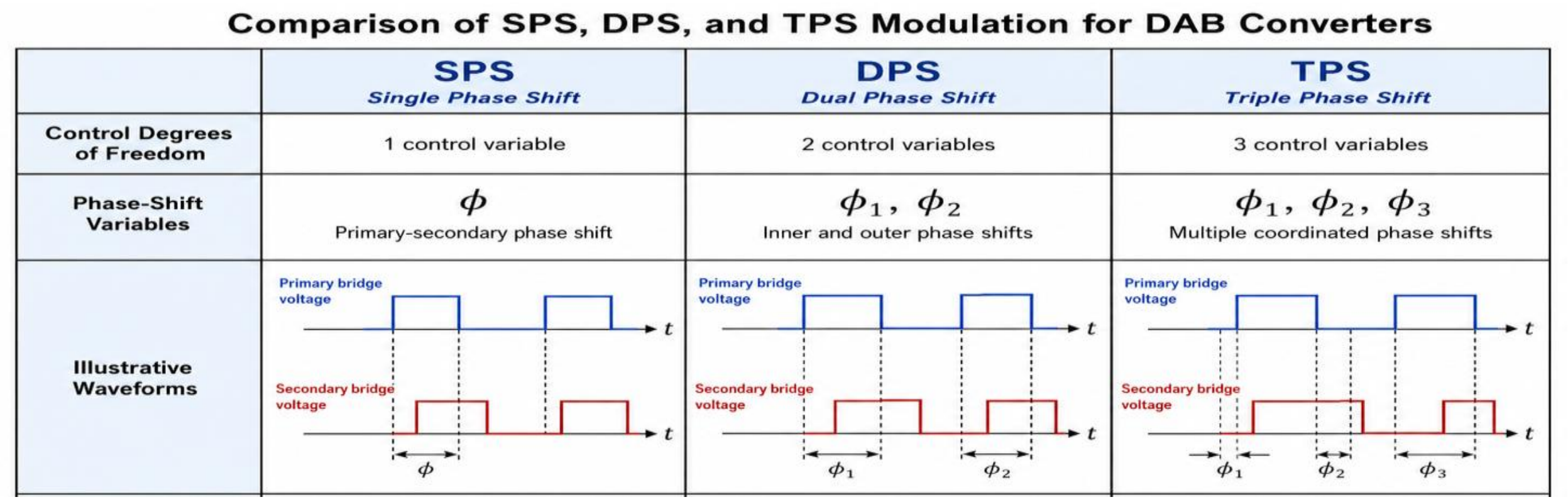

Fig.1: A figure generated by AI used in an undergraduate thesis to elaborate phase shift modulation strategies of the dual active bridge converters.

This figure appeared in the literature-review section of an undergraduate thesis. The student intended to illustrate the differences among several modulation strategies for the dual-active-bridge converter and therefore provided related research papers to GenAI and asked it to generate an illustrative figure.

However, the generated waveforms contain fundamental technical mistakes. The phase-shift variable is incorrectly represented as the duty cycle. In DAB modulation, the phase shift should represent the relative temporal displacement between the corresponding bridge-voltage waveforms.

Compared with its visual presentation, the technical errors in the figure are far more serious and unacceptable. The student accepted the generated figure either because they did not fully understand the underlying modulation concepts or because they took the AI output for granted without sufficient verification. Students must therefore learn not to accept AI output at face value but to question it and verify it against the underlying power electronics concepts.

Taken together, these three examples show that AI-generated technical errors can be difficult for students to detect without sufficient foundational skills and knowledge, a problem that has also been widely discussed in AI research [4], [5]. Similar fabricated information has also been observed in AI-generated references. Educators warn that students should not use Gen AI tools for academic writing unless they are properly trained to obtain critical AI literacy to judge the accuracy of AI output [4].

## III. When Polished Writing Becomes Harder to Read

It has been observed that papers published in *IEEE Transactions on Power Electronics* have become noticeably longer in pages in recent years. According to our sample survey, from 2020 to 2026, the average paper length increased from 13.7 to 14.6 pages, while the average length of the top 20 most-cited papers increased from 13.1 to 14.1 pages. Although this growth cannot be attributed entirely to GenAI usage, the growing use of GenAI has indeed raised concerns about declining readability.

| **Year** | **Paper Average Length** | **Top 20 Most-Cited Papers** |
|---|---|---|
| 2020 | 13.7 pages | 13.1 pages |
| 2026 | 14.6 pages | 14.1 pages |

Table 1. Average paper length in *IEEE Transactions on Power Electronics* in 2020 and 2026.

Readability refers broadly to the ease with which readers can understand a text. For technical writing, readability does not mean avoiding technical depth or specialized terminology. Rather, it means communicating technically complex ideas concisely and without unnecessary linguistic complexity. Although GenAI can avoid most grammatical errors, recent studies have reported increasing syntactic and lexical complexity and declining semantic clarity in AI-assisted scientific writing [6], [7]. In the samples examined in this article, three major categories of readability problems were identified: word and sentence length, lexical and syntactic complexity, and lack of clarity.

### A. Word and Sentence Length

Readability is influenced by both lexical and syntactic features of a text, particularly word length and sentence length. In this study, we adopted two widely used measures of readability: Flesch Reading Ease and Flesch-Kincaid Grade Level [8]. Both Flesch Reading Ease and the Flesch–Kincaid grade level is calculated from the same two inputs, average syllables per word and average sentence length, but each formula assigns those inputs different weights. Higher Flesch Reading Ease scores indicate easier reading, and the Flesch-Kincaid Grade Level estimates the U.S. school grade level typically required to understand a text. For example, a grade level of 12 corresponds approximately to 12th-grade reading, whereas scores of 13–16 generally indicate college-level reading difficulty. Although these measures are not absolute indicators of technical writing quality, they provide useful references for comparing the readability of different texts.

Word length is an important indicator of lexical complexity and readability. It can be measured by the syllable-to-word ratio and the average number of characters per word. Shorter words are usually more concrete and emphatic, reducing the distance between writer and reader, while multisyllable words are often general and abstract and require greater processing effort [9]. Sentence length is another important factor. Although technical writing inevitably involves complex ideas, technical content does not require unnecessarily complicated sentence structures. AI-assisted writing can combine multiple clauses or concepts into a single grammatically polished but lengthy sentence, increasing the cognitive effort required from the reader.

Using these metrics, we compared three expert-written papers published before the GenAI era with three recent student writing samples with AI assistance. As shown in Table II, the expert-written papers consistently employed shorter words, with syllable-to-word ratios ranging from 1.73 to 1.80 and average word lengths ranging from 5.0 to 5.3 characters, compared with 1.97–2.06 syllables per word and 5.6–6.0 characters per word in the student samples. The expert-written papers also had shorter average sentence lengths (16.1 words) than the student samples (16.9 words). Overall, the student samples exhibited significantly lower Flesch Reading Ease scores and higher Flesch-Kincaid Grade Levels, suggesting greater reading difficulty.

Interestingly, although Expert 2 had the longest average sentence length among the expert-written papers (20.3 words per sentence), its relatively short words (1.75 syllables and 5.0 characters per word) have partially offset the effect of sentence length on readability. With a Flesch Reading Ease score of 38.1, Expert 2 remained substantially more readable than the student samples, whose scores ranged from 16.1 to 22.3.

| Text | Avg. Syllables/Word | Avg. Word Length | Avg. Sentence Length | Flesch Reading Ease | Flesch-Kincaid Grade |
|---|---|---|---|---|---|
| Expert 1 [10] | 1.80 | 5.3 | 13.1 | 40.5 | 10.9 |
| Expert 2 [11] | 1.75 | 5.0 | 20.3 | 38.1 | 13.0 |
| Expert 3 [12] | 1.73 | 5.0 | 14.8 | 45.7 | 10.6 |
| Sample 1 (Student Journal paper, accepted) | 1.97 | 5.6 | 20.2 | 19.4 | 15.6 |
| Sample 2 (Student conference paper, accepted) | 2.00 | 5.9 | 15.0 | 22.3 | 13.8 |
| Sample 3 (Student undergraduate thesis) | 2.06 | 6.0 | 15.4 | 16.1 | 14.8 |

Table 2. Readability metrics of expert-written papers and AI-assisted student writing samples.

**B. Lexical and Syntactic Complexity**

Beyond the quantitative measures of word and sentence length, qualitative features of lexical and syntactic complexity have been observed in AI-assisted student writing. The first lexical feature is the use of unnecessarily complicated and abstract words and expressions, also known as professional obfuscation [7]. In technical writing, it is normal to adopt industry-standard terms. The problem arises when linguistic sophistication is added without a gain in technical precision. In such cases, readers spend more effort interpreting the wording.

**Sample 4** ***(Excerpt from a student research paper)***

> *"The topology-independent part of the method is the requirement-to-check-to-decision loop."*

The expression *"requirement-to-check-to-decision loop"* is an AI coinage rather than an established term. Linguistically, it has low semantic transparency [13]. In other words, its meaning cannot easily be inferred from its parts: the reader must deconstruct the noun sequence and infer how *requirement*, *check*, and *decision* are related. This increases lexical processing load without adding technical precision. The underlying idea is simply an engineering design workflow, which can be expressed more directly.

A second readability issue is in the syntactic complexity: the overuse of long strings of nouns or noun phrases. Such structures can make sentences appear compact, but the listed items are not always logically or technically parallel. The following examples, taken from an undergraduate thesis and a student research paper, illustrate this issue.

**Sample 5** ***(Excerpt from an undergraduate thesis)***

> *"The design of a power converter is still a specialized engineering task because electrical specifications, topology choice, component ratings, magnetic design, control assumptions, losses, and safety margins are coupled."*

***Sample 6 (Excerpt from a student research paper)***

> *"A generated design may contain plausible parameters and even a successful simulation result, while component ratings, magnetic design, voltage/current margins, topology suitability, or the limits of the simulation model remain unchecked."*

The grammar is imperfect in Sample 5 in the sense that noun strings in English are expected to convey hierarchy. Sample 6 shows a similar problem. The mixing of parameters and simulation results within the phrase "may contain," and the use of "unchecked" to mean "untested" at the end could be interpreted as writing mistakes.

### C. Lack of Clarity

A third readability problem is lack of clarity. Unlike sentence length or lexical complexity, clarity is difficult to capture using simple quantitative metrics. Sometimes, some content is not technically wrong in power electronics, but it is still very difficult to understand or confusing based on domain knowledge.

**Sample 7** ***(Excerpt from a student research paper)***

> *"Most first-round failures are not language-generation errors alone; they arise when the simulated converter misses efficiency or ripple targets, when stress margins become too weak, or when formula and component checks expose a boundary that requires redesign rather than another text-only answer."*

The student attempted to explain why PE-GPT [14] (a power-electronics design agent) may fail. The intended message is that failures mainly occur when the generated design does not satisfy the expected performance objectives or design constraints. The statement cannot be claimed to be technically incorrect, although "stress margins become too weak" is questionable, but it remains shallow and loosely structured.

First, the sentence relies on a few specific examples for performance objectives, such as efficiency and ripple, to represent the broader concept. Though valid, these examples do not form a structured analysis, which is a typical feature of AI-generated output. As a result, the discussion reads more like an AI-generated aggregation of loosely related examples.

Second, the phrase "formula and component checks expose a boundary" lacks technical clarity. The term *boundary* does not specify what constraint has been reached in the context of power electronics design. Depending on the domain context, it could refer to a performance constraint, a device or component operating limit and so on. In a research paper, such lack of clarity can lead to ineffective communication between authors and readers.

## IV. Implications and Prevention

These examples from power electronics student work demonstrate AI's potential for negative impact on academic integrity and textual readability. Analysis of the sample work points to the key cause behind it: students' lack of ability to evaluate the AI-generated output. Researchers have warned that over reliance on AI in tasks such as academic writing may have long-term costs, including reduced critical thinking and creativity [15], [16], [17]. Recognizing the epistemic risks of inappropriate AI use, educators and students must first acknowledge that there are no shortcuts to developing fundamental competencies in writing and critical thinking. Writing judgment must be developed through sustained reading and practice. Reading high-quality papers helps students build an intuition for effective written communication.

Pedagogically, engineering curricula should evolve to address clear and responsible technical communication in the era of AI. In power electronics technical courses, students can be given AI-generated explanations or figures containing technical errors and asked to identify and correct these errors using their domain knowledge [18]. Such exercises can help students develop critical thinking and reinforce their understanding of fundamental concepts. In technical writing courses, instructors can ask students to prepare an initial draft independently and then use GenAI to revise it. By comparing the two versions, students can reflect on whether the AI-assisted revisions improve clarity and readability or introduce unnecessary complexity and technical inaccuracies [19]. In addition, established guidelines can be introduced explicitly and incorporated into course instruction to help students understand when, how, and to what extent AI tools can be used responsibly in their academic work. The IEEE Power Electronics Society's publication guidelines provide guidance on responsible AI-assisted writing [20].

An important pedagogical implication is that writing education should be strengthened rather than deprioritized in the era of AI. Writing is often compared to a muscle that requires regular exercise to develop and maintain. Similarly, although AI chatbots can rapidly generate human-like text, students still need writing instruction and practice to strengthen their cognitive "muscles". It is not unusual for institutions and students to prioritize technical courses directly related to students' majors, undervaluing writing courses. Consequently, students may lack the skills to critically evaluate AI-generated text, let alone revise it effectively. This concern is consistent with findings from a comparative study of undergraduates in four countries, which identified students' limited exposure to liberal arts courses, including writing, as one factor contributing to their weaker critical thinking performance in the third year of university [21].

For industry, companies should also adapt their practices to better prepare for AI-native engineers. First, AI literacy can be incorporated into recruitment and onboarding, with hiring assessments evaluating candidates' ability to use AI tools and verify AI-generated outputs [22]. Second, companies should strengthen employee awareness of AI ethics, legal and regulatory requirements, and in aspects such as intellectual property, data confidentiality, and accountability. Third, organizations should establish internal AI governance and auditing mechanisms, with clear procedures for documentation, verification, and decision-making. Such measures can help companies leverage younger engineers' AI fluency while mitigating associated risks.

## V. Conclusion

As educators in power electronics and academic writing, we have observed a clear trend: today's AI-native students are increasingly writing with GenAI, and this practice will continue. However, when students lack solid technical knowledge or sufficient language proficiency, they may struggle to recognize technical errors in AI-generated content or notice when readability and clarity are compromised. To raise awareness of these risks, this article examined anonymized examples from undergraduate theses and student research papers in power electronics and identified several recurring problems. Building on the analysis of these challenges and recent research on GenAI in education, we propose pedagogical and institutional implications for students, educators, and industry partners.

Ultimately, the goal is not to discourage the use of AI, but to ensure that its growing adoption in power electronics is accompanied by technical expertise and professional responsibility.

**Acknowledgements**

The authors sincerely thank Dr. Stephanie Watts Butler and Prof. Philip Krein for their constructive feedback on this article.

## **Author Bio** (Contributed Equally)

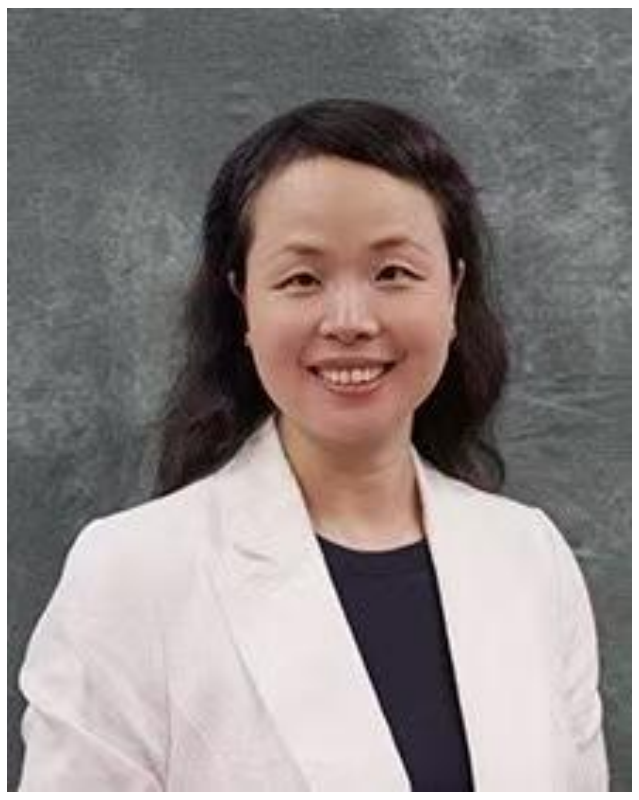

**Min Yang** earned her PhD in Technical Communication and Rhetoric from the Department of English, Texas Tech University, where she also received her second MA in Applied Linguistics. She is a teaching professor of rhetoric and technical writing at the Zhejiang University–University of Illinois Urbana-Champaign Institute. Her research interests include composition pedagogy and theory, feminist rhetorics, and intercultural communication.

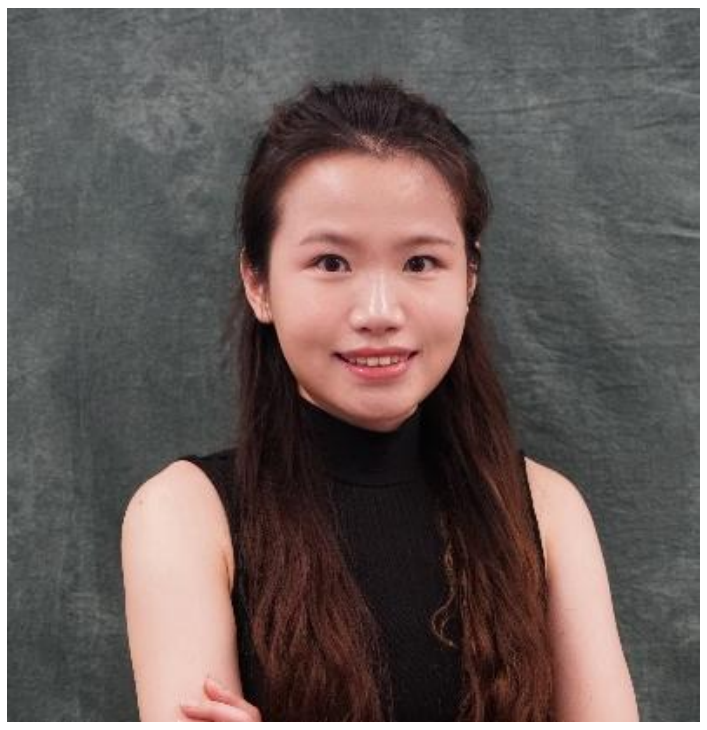

**Fanfan Lin** received the Ph.D. degree in electrical engineering from Nanyang Technological University, Singapore and Denmark Technical University. She is currently an Assistant Professor with the Zhejiang University–University of Illinois Urbana-Champaign Institute. Her research interests include agentic AI and quantum computing applications for power electronics and power engineering. During her undergraduate studies, she minored in English language and literature and conducted research on Chinese–English translation in electrical engineering contexts.